# A Deep Learning System That Generates Quantitative CT Reports for Diagnosing Pulmonary Tuberculosis

Wei Wu[1], Xukun Li[2], Peng Du[2], Guanjing Lang[1], Min Xu[1], Kaijin Xu[1], Lanjuan Li[1]

*1. State Key Laboratory for Diagnosis and Treatment of Infectious Diseases, Zhejiang University, 79 QingChun Road, Hangzhou, Zhejiang 310003;*

*2. Artificial Intelligence Lab, Zhejiang R&D Center of China Telecom Co., Ltd., People's Republic of China.*

Correspondence to Lanjuan Li, PhD, MD, State Key Laboratory for Diagnosis and Treatment of Infectious Disease, Zhejiang University, 79 QingChun Road, Hangzhou, Zhejiang 310003, People's Republic of China

Tel: +86 571 87236458; Fax: +86 571 87236459

E-mail: ljli@zju.edu.cn

## Abstract

We developed a deep learning model-based system to automatically generate a quantitative Computed Tomography (CT) diagnostic report for Pulmonary Tuberculosis (PTB) cases.

501 CT imaging datasets from 223 patients with active PTB were collected, and another 501 cases from a healthy population served as negative samples. 2884 lesions of PTB were carefully labeled and classified manually by professional radiologists.

Three state-of-the-art 3D convolution neural network (CNN) models were trained and evaluated in the inspection of PTB CT images. Transfer learning method was also utilized during this process. The best model was selected to annotate the spatial location of lesions and classify them into miliary, infiltrative, caseous, tuberculoma and cavitary types simultaneously. Then the Noisy-Or Bayesian function was used to generate an overall infection probability. Finally, a quantitative diagnostic report was exported.

The results showed that the recall and precision rates, from the perspective of a single lesion region of PTB, were 85.9% and 89.2% respectively. The overall recall and precision rates, from the perspective of one PTB case, were 98.7% and 93.7%, respectively. Moreover, the precision rate of the PTB lesion type classification was 90.9%.The new method might serve as an effective reference for decision making by clinical doctors.

## 1. Introduction

Pulmonary tuberculosis (PTB) is one of the leading respiratory infectious diseases monitored worldwide, as reported by the World Health Organization.[1] At present, China is still one of the 22 countries with a high PTB burden worldwide. The number of patients with PTB in China ranks third in the world, barely after India and Indonesia.[2,3] In China, the number of reported cases of PTB ranked the second-highest infectious disease, after viral hepatitis.[4] Therefore, correct detection and diagnosis of PTB are very crucial.

Deep learning[5] has been gradually applied to computer-aided diagnosis (CAD) with the rapid development of big data and artificial intelligence (AI) in recent years.[6-8] Relevant studies were conducted on the diagnosis of pulmonary nodules and lung cancer worldwide.[9-11] Numerous companies have released intelligent diagnostic systems for lung nodule detection, such as the Dr. Watson system from IBM. At the same time, some well-known academic institutions and organizations also launched competitions for lung nodule detection on computed tomography (CT) images. Of these, the most famous were the Lung Nodule Analysis 2016 (LUNA16)[12] and the Data Science Bowl 2017 (DSB2017), which were held by the notable data science website Kaggle. These open-sourced datasets had incubated a series of excellent detection and segment algorithms.

However, few recent studies have explored the detection and classification of PTB

infection. The progress in this field has been relatively slow compared with the lung nodule domain because fewer open-sourced CT image datasets of PTB are available. Moreover, the much wider distribution and different characteristics of PTB lesion regions compared with those of lung nodules have also made it difficult to investigate. Despite the differences in morphological features between PTB lesion and pulmonary nodule, some of the open-sourced intelligent detection methods for pulmonary nodule still have a considerable reference value for PTB detection, for example, data preprocessing and image segmentation.

The traditional CT image processing methods are used mainly in the preprocessing stage, while the image segmentation and classification process mostly take advantage of certain deep learning algorithms.[13-15] Setio et al.[16] adopted a convolutional neural network (CNN) to detect pulmonary nodules by extracting features of 2D images. They found that the detection sensitive accuracy of this method was more than 85.4%. Ciompi et al.[17] constructed a labeling system to automatically classify the morphological characteristics of pulmonary nodules into solid, sub-solid, calcified, and nonsolid lesions. Wang et al.[18] proposed a multi-view CNN that integrated several branches of CNN with a fully connected layer, which discriminated various pulmonary nodules more effectively. Zhu et al.[19] adopted 3D Faster R-CNN to detect pulmonary nodules. They used a 3D Dual Path Network to classify the detected pulmonary nodules into benign or malignant and achieved results comparable to doctors' diagnosis. Julian de Wit et al.,[20] who won the second place in the DSB2017 competition, constructed a pulmonary nodule detector through a 3D CNN to predict the possibility of cancer.

In this study, three fine-tuned 3D CNN models were evaluated. The best model was used to detect and classify the PTB lesion regions based on CT image datasets. Moreover, the spatial location of each lesion, the confidence (infection probability) of each single infection, presence of calcifications, classification of each lesion type, overall infection probability, and effective volume of the left and right lungs were digitally achieved according to the output of the AI network model. These reports generated a quantitative evaluation of a single infection region and the whole PTB case, thus greatly assisting clinical doctors to make more accurate diagnostic decisions.

## 2. Method

### Process

Figure 1 shows the whole process of PTB diagnostic report generation in this study. First, the CT images were preprocessed to extract effective lung regions. Second, 3D CNN model were used to segment and classify the lesion region at the same time, and then the overall infection probability was calculated using the Noisy-Or Bayesian function. Finally, a quantitative diagnostic report together with the corresponding labeled CT images was exported for reference.

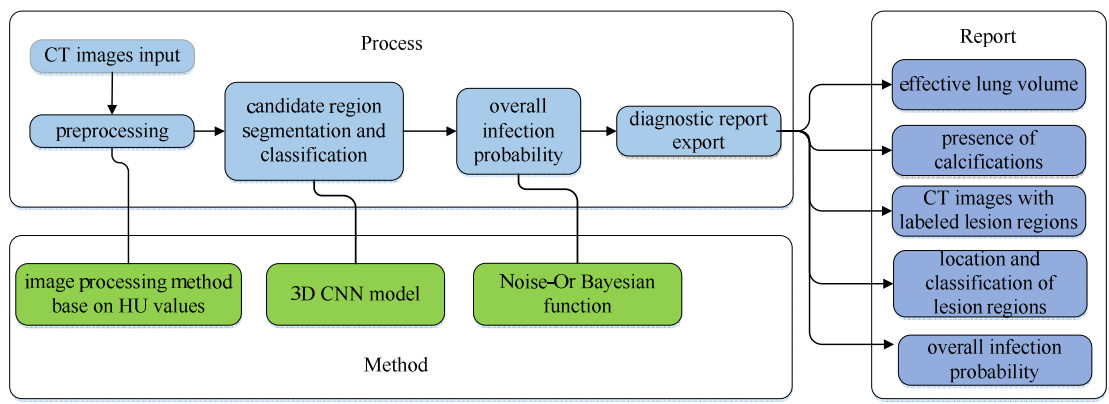

**Figure 1.** Process flow chart. The CT image dataset was first preprocessed according to the Hounsfield Unit (HU) values. The image data were normalized, and the irrelevant part of the image was removed. Then, the image data matrix was fed into the CNN to complete the segmentation and classification of the lesion region at the same time with 3D CNN models. Next, Noisy-Or Bayesian function was used to calculate the overall infection probability from each single lesion region. The final quantitative report included the effective lung volume, presence of calcifications, location and classification of lesion regions, and overall infection probability of this case, along with CT images with labeled lesion regions.

*Dataset introduction*

Five types of active PTB lesions were defined according to the Expert Consensus of Chinese Society of Radiology:[21] miliary, infiltrative, caseous, tuberculoma, and cavitary. (Fig. 2).

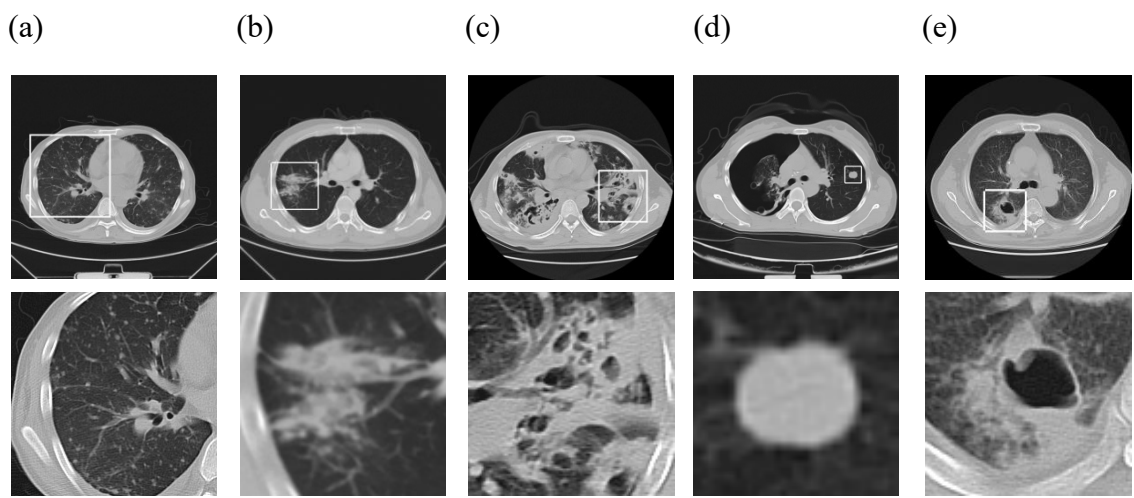

**Figure 2.** Five PTB types: (a) miliary; (b) infiltrative; (c) caseous; (d) tuberculoma; and (e) cavitary. Top: the whole slices. Bottom: the zoomed-in regions.

This study used 501 CT imaging datasets from 223 patients, diagnosed with active PTB at the inpatient department of tuberculosis of the Affiliated Hospital of Zhejiang University from 2016 to 2018. Moreover, another 501 CT image datasets from a healthy population in the same hospital were added as negative samples. Therefore, 1002 CT imaging datasets, all in Digital Imaging and Communications in Medicine (DICOM) format, were used. The ethics committee of the First Affiliated Hospital, College of Medicine, Zhejiang University approved this study. All participants or their legal guardians signed the informed consent form before the study.

The lesions of PTB were manually annotated by two professional radiologists. Together, 2884 (117 miliary, 2255 infiltrative, 135 caseous, 91 tuberculoma, and 286 cavitary) regions were labeled as the PTB lesion (5.8 lesions per CT case on average). The whole cohort were divided into 752 cases of the training set (75%), 100 cases of the validation set (10%), and 150 cases of the test set (15%). Only patients with a single CT case were selected for the test set to avoid the testing deviation that the cases of the same patient had been used during training. Moreover, all these three datasets comprised the same number of CT cases from patients and healthy persons.

***Dataset preprocessing***

To facilitate the detection of PTB lesions, the CT images were resampled to keep the voxel of CT image to $1 \times 1 \times 1$ mm$^3$ measured in the real space following the rule of nearest-neighbor interpolation. Then, the resampled CT sets were preprocessed to generate masks of the effective lung region so as to eliminate the unrelated regions before the training of the deep learning model.

1.  As the digital gray scale image had the pixel value ranging [0, 255], the resampled CT raw data were converted from the Hounsfield Unit (HU) to the aforementioned values interval accordingly. The HU data matrix was clipped within [－1200, 600] (any value beyond this was set to －1200 or 600 accordingly) and then linearly normalized to [0, 255] to fit into the digital image format as shown in Figure 3a.

2.  A fixed threshold (﹣600) was used to binarize the resampled CT images, and bones and soft tissues such as blood vessels and muscles with substantial HU values were filtered out (Fig. 3b).

3.  All connected components smaller than 0.3 $cm^2$ and having eccentricity larger than 0.99 were removed to eliminate some high-luminance radial imaging noise. The components (usually clothes and accessories other than the human body) with the distance to the center of the CT image more than 6.2 cm were also removed. Furthermore, the components with volume between 450 and 7500 $cm^3$ were remained, as shown in Figure 3c. The range in the present study was expanded compared with those reported by lung nodule detection studies[22], which ranged from 680 to 7500 $cm^3$. The nodule detection study usually focused on small regions, while lesions could be more massive for PTB cases.

4.  The extracted mask in step 3 was eroded into two sectors and then dilated to the original size to remove small black holes (Fig. 3d).

5.  Convex hull operation was performed on the effective region, which was extracted from the previous step, to include lesion regions attached to the outer wall of the lung (Fig. 3e).

6.   The matrix data of images in step 1 were multiplied by the masks exported from step 5 to obtain the final effective pulmonary region for further processing. The space out of the mask was filled with 170, which was equivalent to 0 if converts back to HU value. (Fig. 3f)

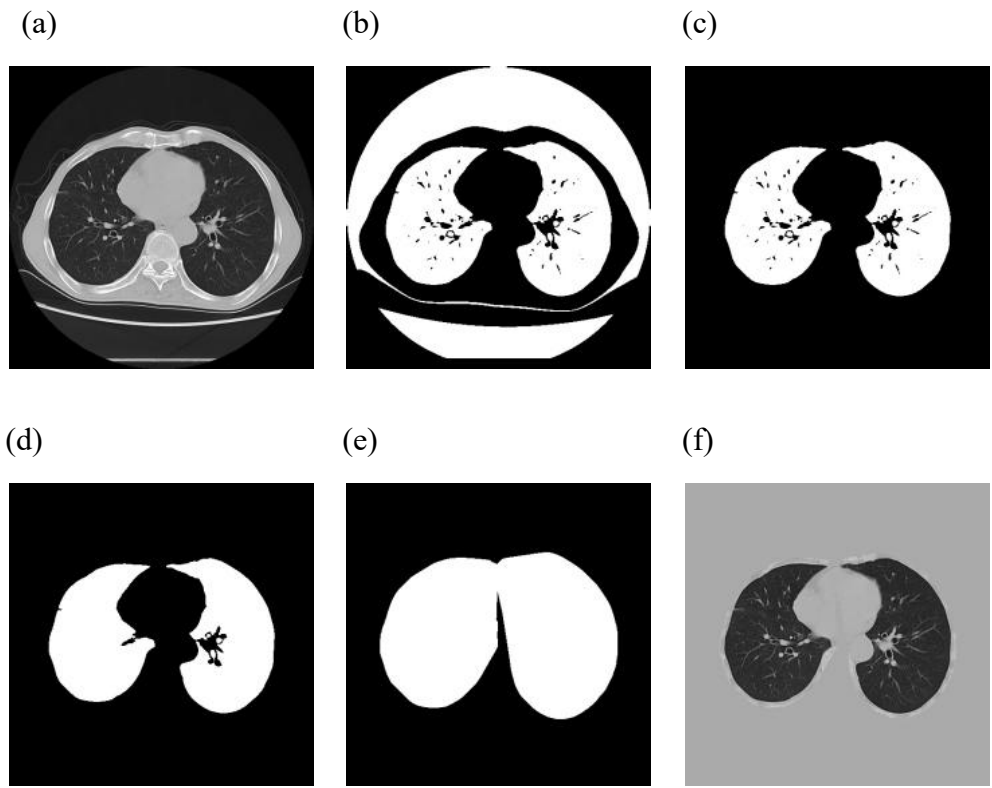

**Figure 3.** Image data preprocessing. (a) Normalized resampled CT image; (b) binarized images with HU equals to -600; (c) removal of unrelated regions; (d) erosion and dilation operation; (e) convex hull operation to create the mask; and (f) normalized CT image multiplied by the mask to generate a valid pulmonary region.

***PTB data process and augment***

To reduce the influence of the uneven distribution of different PTB lesion types in the present dataset, types with fewer specimens were expanded correspondingly. The sampling possibility of miliary, caseous, and tuberculoma cases was expanded 10 times, and that of cavitary cases was expanded 5 times during the training to balance the specimen number with the infiltrative type, which was the dominated type among all. At the same time, generic data expansion mechanisms such as random clipping and left–right flipping were performed on specimens to increase the number of training samples and prevent data overfitting.[23]

***Deep learning model for segmentation and classification***

*Network structure*

3D u-net[24,25] and v-net[26] are most widely used in the domain of medical image segmentation. They are consisted with two network paths: contracting and expanding paths. The images are firstly fed into the contracting path to finish the down-sampling process to capture the context information. Then the up-sampling process is completed in the symmetrical expanding path to obtain precise localization information of the images. At the same time, the feature maps with the equal dimension from both paths are concatenated together, which facilitates to keep the detailed of the neural network from the contracting path during data propagation.

The output layer of 3D u-net and v-net had to be replaced as they were originally designed to generate the segmentations of images, while the target of this study was the location and classification of the TB lesion regions. In our implementation, all three evaluation network models were consisted with two parts: feature extraction part and Region Proposal Network (RPN)[27] output layer. The feature extraction part generated feather maps and enabled the network to capture multi-scale information. The output format of RPN enabled the network to generate proposals (the bounding boxes of the predicted lesions) and classification directly.

Three 3D CNN network models, with different feature extraction structures followed by the same RPN output layer, were designed and evaluated.

1.  The first model (referred as 3DUNET-RPN) used 3D u-net backbone as the feature extraction part, as shown in Figure 4.

2.  The second model (referred as VNET-RPN) used v-net backbone as the feature extraction part, as shown in Figure 5.

3.  The third model (referred as VNET-IR-RPN) used a modified v-net backbone with inception-resnet[28] block added, as shown in Figure 6. The inception-resnet block was consisted with convolution operations with different kernel sizes ($1 \times 1 \times 1$, $3 \times 3 \times 3$), which could extract more detailed features from various receptive fields. Concatenate operation was used to amalgamate multi-scale features to enhance the feature extraction procedure.

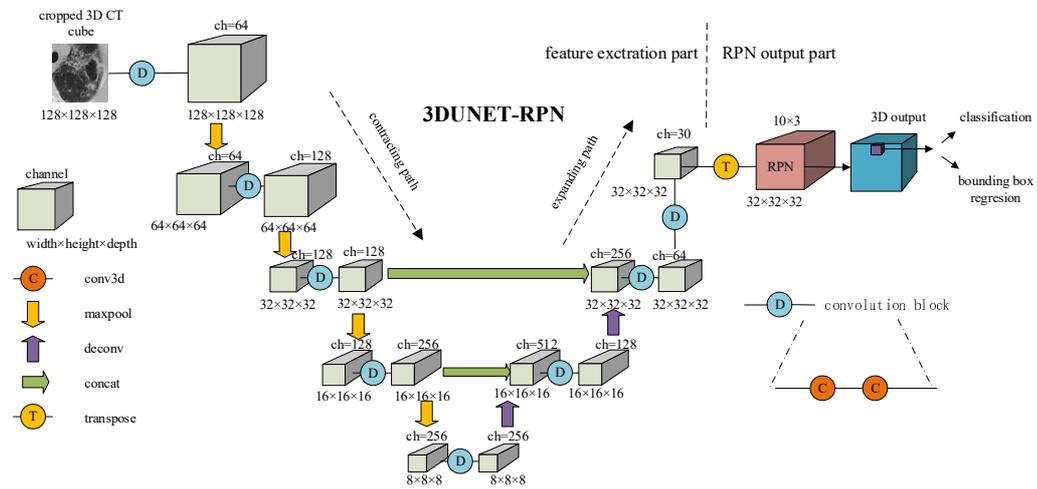

**Figure 4.** 3DUNET-RPN network structure;

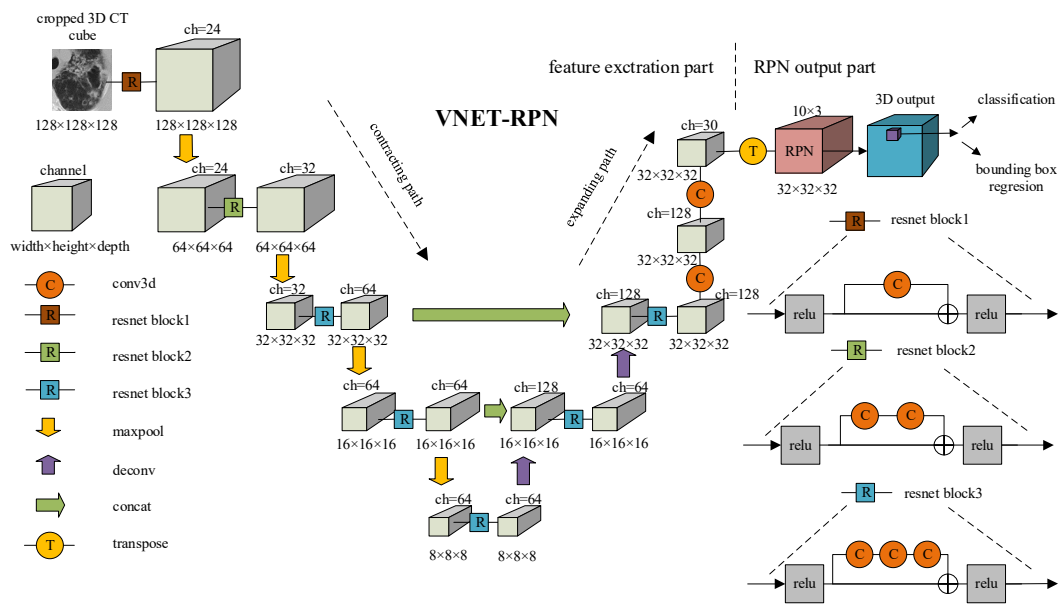

**Figure 5.** VNET-RPN network structure.

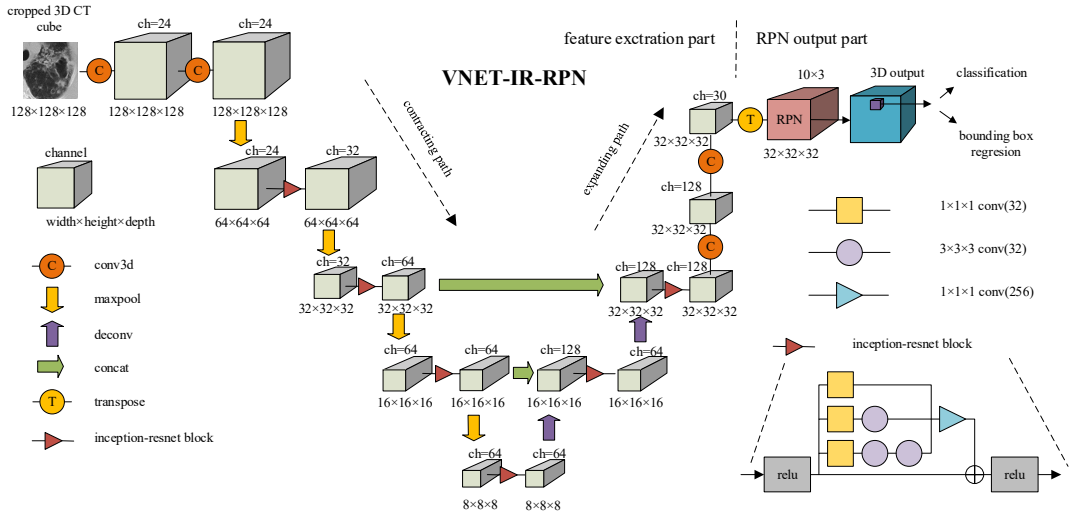

**Figure 6.** VNET-IR-RPN network structure.

*Definition of the loss function*

The total loss $L_{total}$ function included confidence loss $L_{conf}$, location regression loss $L_{reg}$, and classification loss $L_{class}$. The first two losses were the nodule detection loss function,[22] while the classification loss was added for PTB study.

The confidence loss $L_{conf}$ is a cross entropy loss to measure whether this proposal was a valid target:

$$L_{conf} = -(p\log(\hat{p}) + (1-p)\log(1-\hat{p}))$$

(1)

where $p$ is the ground truth and $\hat{p}$ is the predicted value.

The ground truth bounding box of a PTB lesion is denoted by ($G_x$, $G_y$, $G_z$, $G_d$) and the bounding box of an anchor is denoted by ($A_x$, $A_y$, $A_z$, $A_d$), where the first three elements stand for the center point and the last element for the side length.

The regression labels of the bounding box included the regression of the center point $(d_x, d_y, d_z)$ and the side length $d_d$

$$d_x = (G_x - A_x) / A_d$$
$$d_y = (G_y - A_y) / A_d$$
$$d_z = (G_z - A_z) / A_d \qquad (2)$$
$$d_d = \log(G_d / A_d)$$

The corresponding predictions were $(\hat{d}_x, \hat{d}_y, \hat{d}_z, \hat{d}_d)$. Then, the total regression location loss is defined by

$$L_{reg} = \sum_{k \in \{x,y,z,d\}} S(d_k, \hat{d}_k) \qquad (3)$$

where the function $S$ was defined as

$$S(d, \hat{d}) = \begin{cases} |d - \hat{d}|, if \ |d - \hat{d}| > 1 \\ (d - \hat{d})^2, else. \end{cases} \qquad (4)$$

$L_{class}$ is the cross entropy loss of five-classification dimension,

$$L_{class} = -\sum_i y_i \log \hat{y}_i \qquad (5)$$

where $y_i$ is the ground truth label and $\hat{y}_i$ is the predicted label.

Intersection over Union (IoU), which is equal to the overlapped area of the bounding boxes of two objects divided by their united area, is an evaluation metric used to measure the accuracy of an object detector on a particular target and define the tags of each anchor box in the present study. The anchor box with IoU larger than 0.5 was treated as a positive sample ($p = 1$), while that with IoU small than 0.02 was regarded as a negative sample ($p = 0$). Others were neglected during training and validation.

Then, the loss function ($L_{\text{total}}$) is defined as follows:

$$L_{\text{total}} = L_{\text{conf}} + p(L_{\text{reg}} + \lambda L_{\text{class}})$$

(6)

$p$ equals to 1 and 0 when the box is a positive sample and a negative sample respectively; and $\lambda$ is set to 0.5 according to the setting of Yolo,[29] which is a well-tuned deep learning algorithm for 2D object segmentation and identification.

***Training process***

*Patch-based input for training*

The 3D CT image patches were cropped from the lung images and then fed into the network individually. They were randomly selected based on the following rules. First, 70% of the patches contained at least one ground truth PTB lesion, which indicated that either the center point or more than 12 mm margin at each dimension from the region were included. The rest 30% were cropped from the healthy area to ensure the coverage of enough negative samples.

The Clipped 3D CT image patches were cropped from the lung scans to save GPU memories and then fed into the network individually. The size of the patch was 128 × 128 × 128 × 1 (height × length × width × channel). The output of the last convolution network was resized to 32 × 32 × 32 × 3 × 10 in the transpose layer, where the last two dimensions corresponded to the anchors in the RPN network and the location and classifications, respectively. Three different scales of anchors with the side length of 10, 40, and 80 mm were used. Hence, the output layer had $32 \times 32 \times 32 \times 3$ anchor bounding boxes. The 10 regression dimension was $\{p_i, x_i, y_i, z_i, d_i, t_0, t_1, t_2, t_3, t_4\}$ where $p_i$ is the confidence; $x_i$, $y_i$ and $z_i$ denote for the center of the candidate; $d_i$ is the side length of the region, and $t_{0-4}$ is the possibility of 5 PTB types individually.

*Transfer learning*

To accelerate the convergence rate of the PTB analysis models, transfer learning was utilized in the study by first training models for the task of lung nodule detection using two open-sourced pulmonary CT datasets LUNA16 and DSB2017, which contained 888 and 2101 lung CT nodule analysis cases, respectively. The outputs of the nodule detection models included the coordinates of the center point, the side length, and the confidence of the nodule region. Then this pre-trained nodule detection models were used to initialize the network for PTB study.

*PTB training*

In the next PTB training stage, only the output layer and the loss function were modified to be included into the lesion classification task, while the rest of the network structure remained unchanged.

At in the beginning of the training, the PTB analysis network was initialized with the parameters from the pre-trained nodule detection model (as they had the exactly same network structures) except for the output layer, which was initialized randomly with the normal distribution.

***Performance evaluation***

A non-maximum suppression algorithm[30] was first performed on detected PTB lesion regions to remove repeated candidate bounding boxes. If the central coordinate of the remaining box was within the radius of the human annotated lesion region, the result was marked as true positive (TP); otherwise it was false positive (FP). False negative (FN) indicated that no predicted bounding box was corresponding to a human annotated region to measure the number of issues missed by the model. Accordingly, *Recall*, *Precision,* and the more balanced *F*1_*score* were used to measure the performance of the deep learning model:

$$Recall = \frac{TP}{TP + FN} \tag{7}$$

$$Precision = \frac{TP}{TP + FP} \tag{8}$$

$$F1\_score = 2 \times \frac{Precision \times Recall}{Precision + Recall} \tag{9}$$

The test dataset consisted of 150 cases, including 75 cases with PTB lesions and 75 normal cases from 75 healthy people, with 412 valid PTB lesion regions.

***Quantitative diagnostic report***

The exported CT images were converted back to the original size for easy to review. The final quantitative diagnostic report, based on the detection and classification of PTB information, included the overall infection probability, effective volume of the left and right lungs, classification of lesion type, spatial location of infection, and presence of calcifications. The original CT images with corresponding annotated lesion regions were also exported.

1. Overall infection probability of the left and right lungs

According to the confidence level of each single detected lesion, the overall infection level ($P$) of the left and right lungs was calculated using the probability formula of the Noisy-Or Bayesian function[31] as follows:

$$P = 1 - \prod_i (1 - P_i) \tag{10}$$

where $P_i$ represents the infection possibility of the $i$th lesion in this single lung.

2. Effective volume of the left and right lungs

The effective volume of lungs had consulting value for doctors in medical diagnosis.[32,33] The effective volume of a single lung was calculated by extracting the effective region in the original CT images according to the value of HU (threshold equals to －600). By removing the blood vessels, soft tissues, and lesion regions, the volume $V_i$ of a single CT image was calculated as $V_i = S_i \times h$, where $S_i$ is the effective lung region of the $i$th piece and $h$ is the physical thickness between two adjacent slices. Then, the total volume (in real physical size) of the effective lung was measured as follows:

$$V_{\text{total}} = \sum_{i=0}^{k} S_i \times h$$

(11)

3. Spatial location of infection

As a 3D system, the number of CT image slices was used instead of coordinate $z$. The parameters $x$, $y$, and $d$ (in pixel size) represented the center point and the side length of this single lesion region. The origin of coordinates of $x$ and $y$ was at the lower-left corner of each CT image.

4. Classification of lesion type

The segmented PTB lesion regions were classified into (1) miliary, (2) infiltrative, (3) caseous, (4) tuberculoma, and (5) cavitary types.

5. Recognition of the presence of calcification

According to clinical experience, a HU value of more than 120 of the nodule with an effective region of at least 3 pixels indicated the presence of calcification.

6. CT image annotations

The location of lesions was annotated as the bounding box on the CT slices corresponding to the output of the deep learning model, together with their infection probabilities, types, and presence of calcification. Only the image slice with the center of the lesion was labeled to avoid confusion.

## 3. Results

### *Evaluation platform*

An Intel i7-8700k CPU together with NVIDIA GPU GeForce GTX 1080 was used as the testing server.

### *Training curve*

All three network models were trained with and without pre-trained models seperately (shown in Table 1). The results indicated that without pre-trained models, the models did not converge at all after a few hundreds of epochs. Therefore, the pre-trained models were used for the next steps of this study. As the epoch number of training iterations increased to more than 350, the loss value did not decrease or increase obviously, suggesting that the models converged well to a relative optimal state without distinct overfitting. The training curves of the loss value for each CNN model were shown in Figure 7. Both evidences indicated that VNET-IR-RPN network achieved better performance and convergent rate on training dataset than the other two networks.

| Model | Pre-trained | Convergence | Loss | Detection accuracy | Classification accuracy | #Epoch |
|-------|-------------|-------------|------|--------------------|------------------------|--------|
| 3DUNET-RPN | N | N | — | — | — | 400 |
| 3DUNET-RPN | Y | Y | 0.263 | 89.2% | 91.8% | 350 |
| VNET-RPN | N | N | — | — | — | 400 |
| VNET-RPN | Y | Y | 0.151 | 92.6% | 94.2% | 260 |
| VNET-IR-RPN | N | N | — | — | — | 400 |
| VNET-IR-RPN | Y | Y | 0.134 | 93.6 | 94.8 | 300 |

**Table 1.** All three models did not converge without pre-trained models. The detection and classification accuracy values were based on the training set. The classification accuracy would be calculated only when it was a true positive region.

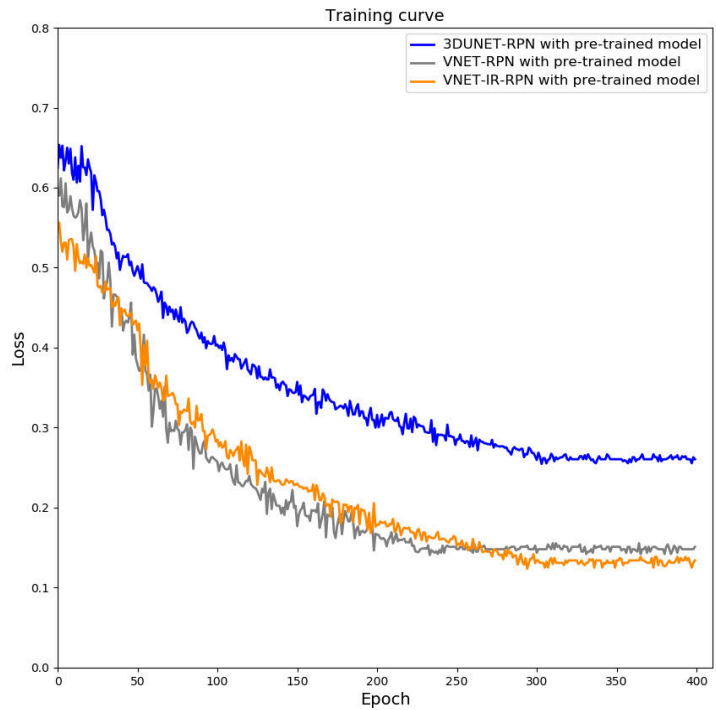

**Figure 7.** Training curve of loss value. The loss referred to the total loss function as listed in Eq. (6).

*Model performance on test dataset*

The performance of all three 3D CNN models were evaluated on the test set, which consisted of 150 cases, including 75 cases from PTB group and 75 normal cases from healthy group, with 412 valid PTB lesion regions. The detection accuracy was evaluated firstly, as the classification accuracy would be calculated only when it was a true positive region.

The Free Response Operation Characteristic (FROC) analysis was utilized to evaluate the performance of different models on the test dataset, as shown in Figure 8. To facilitate directly quantitative comparisons among models, FROC system score was computed, which was the average of the recall at seven predefined false positive per scan (1/8; 1/4; 1/2; 1; 2; 4; and 8).

The corresponding FROC system score for 3DUNET-RPN, VNET-RPN and VNET-IR-RPN were 0.875, 0.901 and 0.917, which showed that VNET-IR-RPN had the best performance averagely. This result also highlighted the effectiveness and efficiency of inception-resnet blocks in this 3D inspection architecture. Therefore, VNET-IR-RPN model was used for the rest of this study.

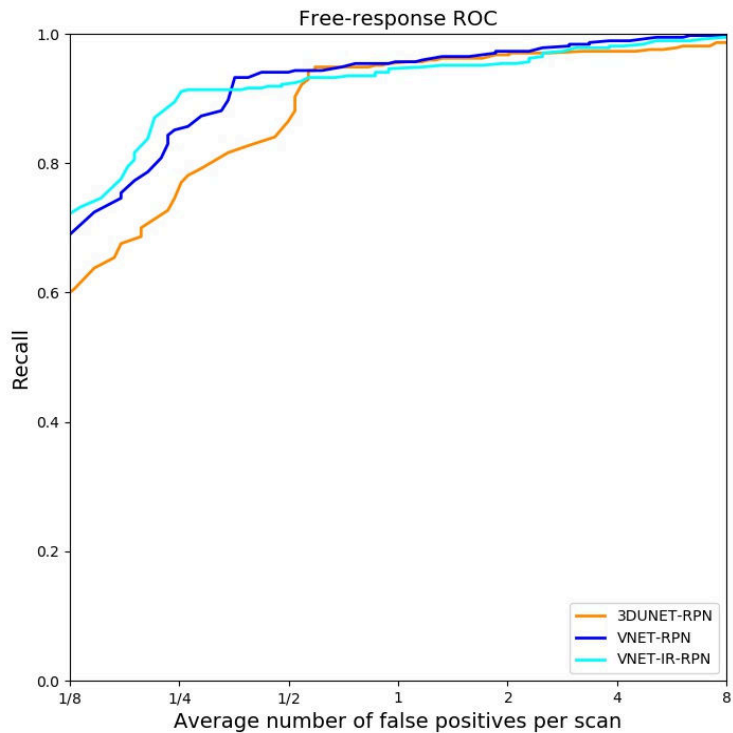

**Figure 8.** The FROC curve of different models.

In order to achieve the maximum value of $F1\_score$, we set the threshold (classified as lesion region if the predicted probability is higher than the threshold) to 0.38. In the test dataset, 397 candidate regions were detected by the VNET-IR-RPN model, including 354 TP and 43 FP. Moreover, 58 regions were observed as FN. The corresponding *Recall*, *Precision*, and $F1\_score$ were 85.9%, 89.2%, and 0.875, respectively.

From the perspective of a whole PTB case, 79 cases (74 cases from PTB group and 5 cases from healthy group) were detected to have at least one lesion, while 71 (1 cases from PTB group and 70 cases from healthy group) cases were left as no findings.

The *Recall, Precision,* and *F1_score* of infected cases were 98.7%, 93.7% and 0.961, and those of the healthy cases were 93.3%, 98.6% and 0.959, respectively.

In addition to the spatial label, the 354 detected TP samples were classified using the VNET-IR-RPN model at the same time with segmentation. The results showed that 322 regions were correctly cataloged, and the classification precision rate was 90.9%. The results of sample detection and classification case are shown in Figure 9.

(a)                                         (b)

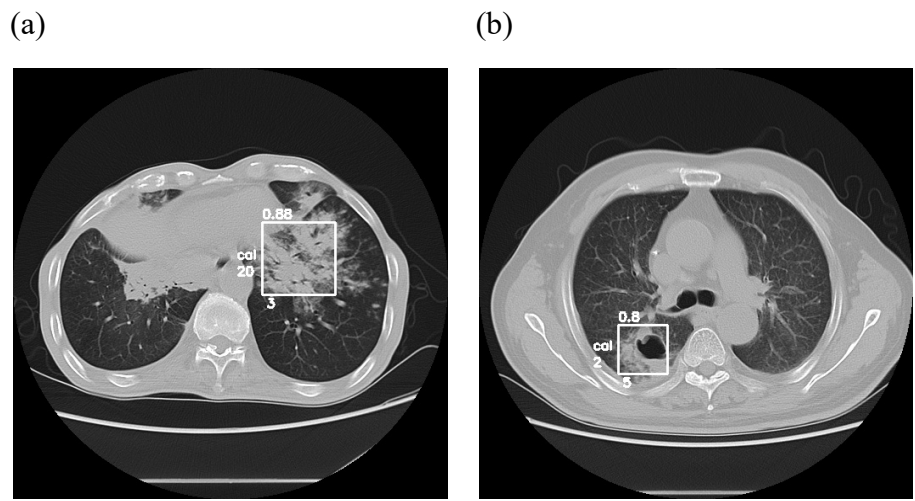

**Figure 9.** Detection and classification results of PTB lesion regions. (a) The digit 0.88 on the top of the bounding box denoted the probability of infection in this lesion region; the digit 3 at the bottom of the box for type 3 (caseous) PTB; and the "cal" and digit 20 on the left of the box for the presence of calcifications with 20 pixels. (b) The digit 0.8 for the probability of infection in this lesion region; the digit 5 at the bottom for type 5 (cavitary) PTB; and the cal and digit 2 on the left for the presence of cal with 2 pixels. (Usually, at least 3 pixels indicated the effective presence of calcification)

### *Example of the diagnostic report*

An example of an exported diagnosis report, consisting of a summarized description report and a series of images with labeled lesions accordingly, is shown in Table 2 and Figure 10.

Quantitative diagnostic report of PTB

Name: xxx, Date of Birth: xxx, Gender: xxx, Study Date: xxx

| Left lung: | Right lung: |
|---|---|
| Overall IP: 98.8% | Overall IP: 98.3% |
| Effective volume: 974.16 (cm$^3$) | Effective volume: 1352.57 (cm$^3$) |
| 24th slice, x = 367, y = 377, d = 35, type: 2 (infiltrative), IP = 75.0%, Cal.: no | 26th slice, x = 164, y = 196, d = 32, type: 2 (infiltrative), IP = 71.0%, Cal.: no |
| 26th slice, x = 400, y = 314, d = 37, type: 2 (infiltrative), IP = 80.0%, Cal.: no | 39th slice, x = 163, y = 315, d = 31, type: 2 (infiltrative), IP = 71.0%, Cal.: no |
| 34th slice, x = 359, y = 383, d = 48, type: 2 (infiltrative), IP = 65.0%, Cal.: no | 39th slice, x = 179, y = 244, d = 26, type: 2 (infiltrative), IP = 65.0%, Cal.: no |
| 45th slice, x = 323, y = 370, d = 59, type: 5 (cavitary), IP = 75.0%, Cal.: yes | 44th slice, x = 147, y = 226, d = 26, type: 2 (infiltrative), IP = 78.0%, Cal.: no |
| | 49th slice, x = 177, y = 325, d = 37, type: 2 (infiltrative), IP = 78.0%, Cal.: no |
| | 50th slice, x = 202, y = 239, d = 34, type: 2 (infiltrative), IP = 73.0%, Cal.: no |

**Table 2.** Example of a diagnostic report. The number of CT image slice was used instead of *z*. The parameters *x*, *y*, and *d* (in pixel size) represented the center point and the side length of the lesion region. Cal., Presence of calcification; IP, infection probability.

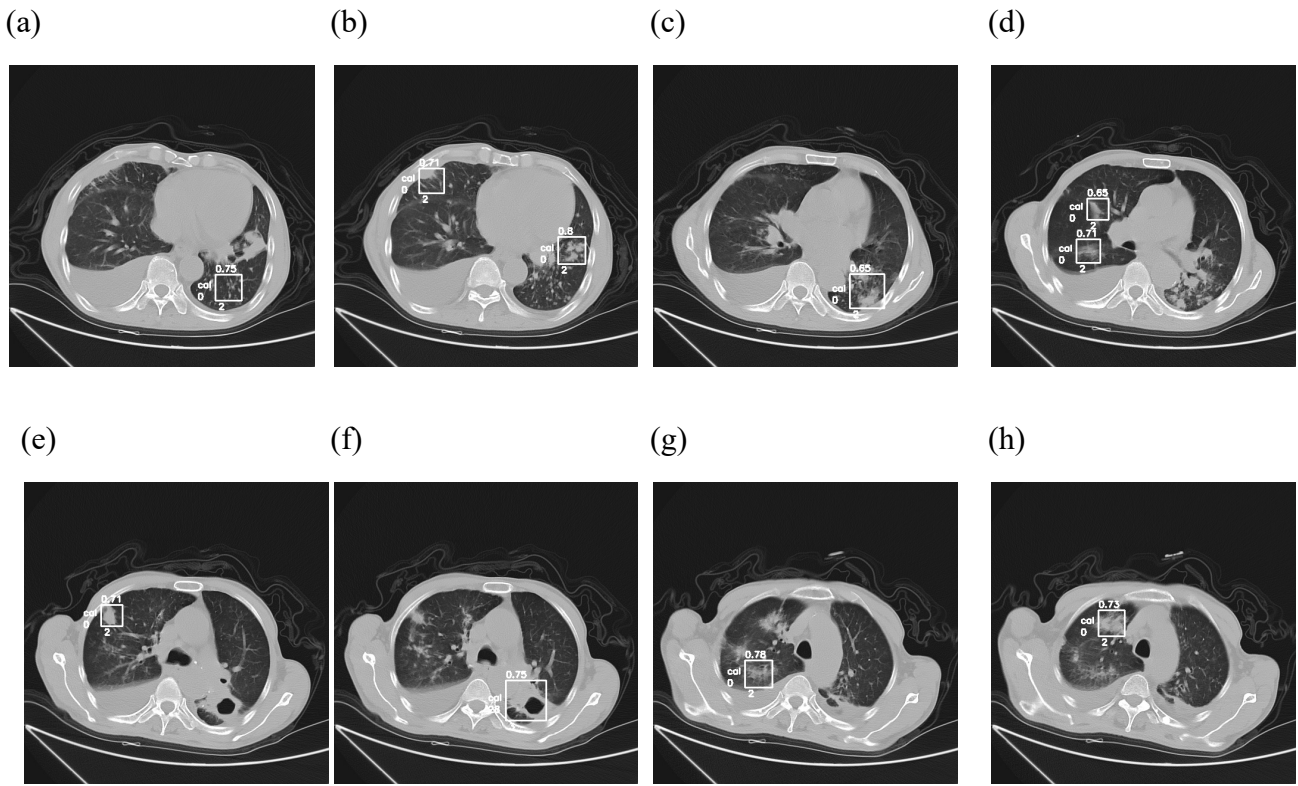

**Figure 10.** Example of a diagnostic result corresponding to the same case as in Table 2 (a) 24th slice; (b) 26th slice; (c) 34th slice; (d) 39th slice; (e) 44th slice; (f) 45th slice; (g) 49th slice;　and (h) 50th slice.

## 4.　Discussion

Currently, most CAD systems focused on PTB using chest x-ray images. CAD4TB is an AI-based CAD system for detecting PTB,[34] which makes judgments based on chest x-ray images as "No TB", "Possible TB", and "Likely TB" scenarios. Lakhani et al.[35] from the Thomas Jefferson University Hospital used a different deep CNN for deep learning and classified chest x-ray images into PTB manifestation or normal with a precision rate of 97.3%. These CAD systems using chest x-ray images could only perform bi-choice or multi-choice operation to detect the presence of TB and might ignore small-scale lesions.

CT imaging is a tomographic imaging method with the characteristics of high resolution and non-invasiveness. A single scan usually generates dozens to hundreds of images to present a complete 3D pulmonary view. Compared with the planar imaging of chest x-ray, CT has more advantages, such as higher spatial precision, clearer display of organs and structures of lesions, and stereoscopic exhibition. However, it is difficult for radiologists to evaluate the severity of a single lesion as well as the whole case accurately on a 3D scale. Controversial decisions can be made by one doctor at different times or by different doctors due to subjective judgments.

In this study, three state-of-the-art 3D deep learning models were adopted to analyze CT images of the lungs, which could use the advantages of 3D CT imaging characteristics to detect various regions and types of lesions effectively. Among them, v-net backbone with inception-resnet blocks achieved the best performance both for the accuracy of detection and classification. Then the exported quantitative report, with overall infection probability, calcification information, lung effective volume, lesions with spatial coordination and corresponding labeled images, may serve as an effective reference for doctors to make decisions.

However, this study had several limitations. First, besides PTB, many other signs existed for intrathoracic TB, including lesions in the pleural cavity, pericardium, bones, and lymph nodes. This study focused only on the five typical types of pulmonary lesions and ignored the other signs. Second, many pulmonary issues other than PTB still existed, such as infectious diseases (bacteria, fungi, viruses, parasites and so on) and non-infectious diseases (tumors and vasculitis and so on). Due to the lack of relevant training samples, other pulmonary lesions could not be correctly identified in this study and were misjudged as one type of PTB. The samples of pulmonary lesions other than TB should be added for effective discrimination in the future. Third, the CT samples in this study were collected from inpatient PTB cases with relative massive region of lesions. The current model might be less sensitive to trivial PTB lesions. Moreover, the proposed model might misjudge some of the lesions due to the false-positive rate; therefore, doctors still needed to review the full CT scan to confirm the result.

Future investigations can be improved from the following aspects. First, for extracting of effective lung regions, a fixed threshold method was used to extract masks in this study. For improvement, more effective segmentation methods can be used in data preprocessing due to the wide distribution and different types of PTB lesion regions. For example, a better pulmonary mask can be achieved by extracting the lung contour by a deep learning regression method. Second, during the full complete TB treatment cycle for one patient, clinical doctors were more concerned about the changes in PTB lesions. Hence, patients needed to be scanned several times. The comparisons should be made before, during, and after the therapy to assess the treatment effect. An artificial intelligence system can be used in the future to analyze all CT cases of one patient along the time sequence with a quantitative comparison of the whole PTB treatment.

## Acknowledgments

This study was supported by the Chinese National Science and Technology Major Project Fund (20182X10101-001).

## Author contributions

Xukun Li and Peng Du developed the network architecture and data/modeling infrastructure, training, testing setup and statistical analysis. Wei Wu, Xukun Li and Peng Du wrote the manuscript. Wei Wu, Kaijin Xu and Lanjuan Li provided clinical expertise and guidance on the study design. Guanjing Lang, Min Xu and Kaijin Xu created the datasets, interpreted the data and defined the clinical labels. Wei Wu and Lanjuan Li initiated the project and provided guidance on the concept and design.

Lanjuan Li supervised the project.

**Competing interests**

The authors declare no competing interests.